\documentclass[final, aps, pra, floatfix, reprint, nofootinbib, citeautoscript, superscriptaddress, balancelastpage, longbibliography, showkeys]{revtex4-2}
\usepackage[T1]{fontenc}
\usepackage[english]{babel}

\usepackage{amsmath}
\usepackage{amssymb}
\usepackage{bm}
\usepackage{wasysym}
\usepackage{textcomp}

\usepackage[normalem]{ulem}

\usepackage{tabularx}
\usepackage{subfigure}
\usepackage{graphicx}
\usepackage[dvipsnames]{xcolor}

\usepackage[colorlinks, urlcolor=blue, citecolor=blue, linkcolor=blue, pdfstartview=FitH]{hyperref}
\usepackage[all]{hypcap}

\def\affiIOFFE{Ioffe\ Institute, 194021 St.~Petersburg, Russia}

\begin{document}
\selectlanguage{english}

\author{M.\ V.\ Petrenko}
\affiliation{\affiIOFFE}

\author{A.\ K.\ Vershovskii}
\email{antver@mail.ioffe.ru}
\affiliation{\affiIOFFE}

\selectlanguage{English}

\title{Anomalous suppression of spin-exchange relaxation in alignment signals in cesium in ultra-weak magnetic fields}
\begin{abstract}
The results of a study of the dynamics of atomic moments alignment in cesium under optical pumping by linearly polarized resonant light in ultra-weak magnetic field are presented. It is shown that there are alignment components whose relaxation does not depend on spin-exchange broadening. The effect of suppression of spin-exchange relaxation in zero magnetic fields is detected, which is similar in its manifestations to the SERF (Spin-Exchange Relaxation Free) effect observed in orientation signals. This observation is interesting from the standpoint of general theory, since the law of conservation of angular momentum responsible for maintaining orientation in the SERF mode should not guarantee the preservation of alignment under the same conditions. A comparison with theoretically calculated parameters of SERF resonances in orientation is given. A qualitative explanation of the observed effect is presented. \end{abstract}


\maketitle

\section{Introduction}\label{sec:1}

Various quantum sensors based on the effects of optical pumping in atoms (primarily hydrogen-like alkalis) and atom-like structures, such as nitrogen-vacancy (NV) color centers in diamond, have become among the most rapidly developing and sought-after devices in applied physics in recent decades. These include atomic clocks  \cite{Kitching_2018,Knappe_Schwindt_Shah_Hollberg_Kitching_Liew_Moreland_2005}, sensors of magnetic field  \cite{Budker_Romalis_2007,Petrenko_Pazgalev_Vershovskii_2021,Petrenko_Pazgalev_Vershovskii_2023}, rotation (gyroscopes)  \cite{Meyer_Larsen_2014,Vershovskii_Litmanovich_Pazgalev_Peshekhonov_2018}, temperature  \cite{Neumann_Jakobi_Dolde_Burk_Reuter_Waldherr_Honert_Wolf_Brunner_Shim_et_al._2013}, electric field strength  \cite{Dolde_Fedder_Doherty_Nobauer_2011}, etc.

The vast majority of alkali atom sensors are based on the optical orientation effect. Orientation is usually understood as both the process and the result of the action of a circularly polarized beam resonant to an optical atomic transition on an ensemble of atoms; in this case, the angular momentum is transferred from photons to the atoms, and a group kinetic and magnetic moment is formed. In the hierarchy of moments, orientation is a first-order moment. It is characterized by an asymmetric distribution of the populations of the Zeeman sublevels, and, as a consequence, a nonzero value of the average moment.

For a long time, the main obstacle to the creation and then increase of sensitivity of optical quantum sensors was the rapid relaxation of oriented atoms during collisions both with the cell walls and with each other (in solid-state sensors – spin-lattice and spin-spin relaxation). It is very important to note that relaxation processes are closely related to conservation laws, primarily – with the law of conservation of angular momentum. Thus, due to this law, spin-exchange processes do not change the total angular momentum of the atomic ensemble. This is why they do not contribute to the relaxation of the longitudinal component of the total moment with respect to the magnetic field – that is, to the longitudinal (determined by time $T_1$) relaxation of orientation. At the same time, these processes are capable of changing the phases of the precession of colliding atoms, while preserving the total moment. This determines the contribution of spin-exchange to the transverse (determined by time $T_2$) relaxation. Note that in classical magnetic resonance described by stationary Bloch equations, the resonance width is determined by the time $T_2$. Therefore, it is transverse relaxation that limits the achievable sensitivity of quantum sensors. This fully applies to SERF sensors based on the level crossing (Hanle) resonance  \cite{Aleksandrov_Bonch-Bruevich_Khodovoi_1967} in zero magnetic field.

Methods for suppressing relaxation on the cell walls were soon found: filling the cell with buffer gases  \cite{Franzen_1959} that slow down the motion of atoms, or applying anti-relaxation coatings \cite{Bouchiat_Brossel_1966}. Later, W. Happer showed  \cite{Appelt_Ben-AmarBaranga_Young_Happer_1999} that spin-exchange relaxation can also be partially or completely suppressed under conditions where most of the colliding atoms are in identical states. This subsequently led to a real breakthrough in quantum magnetometry: SERF zero-field sensors were created  \cite{Ledbetter_Savukov_Acosta_Budker_Romalis_2008, Kominis_Kornack_Allred_Romalis_2003}, and then non-zero-field sensors using ``stretched'' states  \cite{Petrenko_Pazgalev_Vershovskii_2021,Scholtes_Schultze_IJsselsteijn_Woetzel_Meyer_2011,Schultze_Schillig_IJsselsteijn_Scholtes_Woetzel_Stolz_2017}. These methods of suppressing spin-exchange relaxation are also based on conservation of angular momentum: in an ensemble in which all atoms have the same (maximum in modulus) projection of angular momentum and a common phase of precession, a change in this state is impossible without changing the total angular momentum and its projection. In particular, the SERF effect, which ensures almost complete suppression of spin-exchange relaxation, is due to the conservation of the total angular momentum of both hyperfine sublevels, which due to ultra-frequent collisions precess at a certain common frequency \cite{Happer_Tam_1977}.

All these considerations, however, turn out to be useless in the case of alignment, since it is characterized by a zero value of the average moment. Alignment is a second-order moment in the hierarchy. It arises when atoms are pumped by linearly polarized or unpolarized light; it is characterized by a symmetric distribution of the populations of the Zeeman sublevels. Quantum magnetic field sensors based on alignment, despite their smaller distribution, are characterized by certain advantages, such as the absence of dead zones \cite{Ben-Kish_Romalis_2010,Wang_Wu_Xiao_Wang_Peng_Guo_2021}, reduced orientation error  \cite{Hovde_Patton_Versolato_Corsini_Rochester_Budker_2011,Zhang_Kanta_Wickenbrock_Guo_Budker_2023}, small drift, and accuracy over long observation times  \cite{Rosner_Beck_Fierlinger_Filter_Klau_Kuchler_Rosner_Sturm_Wurm_Sun_2022}. It is interesting to note that sensors based on NV centers use a mechanism of pumping by unpolarized radiation, which creates alignment in the NV ensemble.

A theoretical description of the alignment effects in alkali metals for stationary cases is given in  \cite{Akbar_Kozbial_Elson_Meraki_Kolodynski_Jensen_2024,Weis_Bison_Pazgalev_2006}, and developed in detail for quasi-stationary cases in  \cite{Meraki_Elson_Ho_Akbar_Kozbial_Kolodynski_Jensen_2023}. The idea of a zero-field magnetometer based on the alignment effect, which is attractive due to the extreme simplicity of the optical scheme, is also developed in  \cite{LeGal_Lieb_Beato_Jager_Gilles_Palacios-Laloy_2019,Breschi_Weis_2012,Breschi_Grujic_Weis_2014}.

\begin{figure*}[!t]  
	\includegraphics[width=\linewidth]{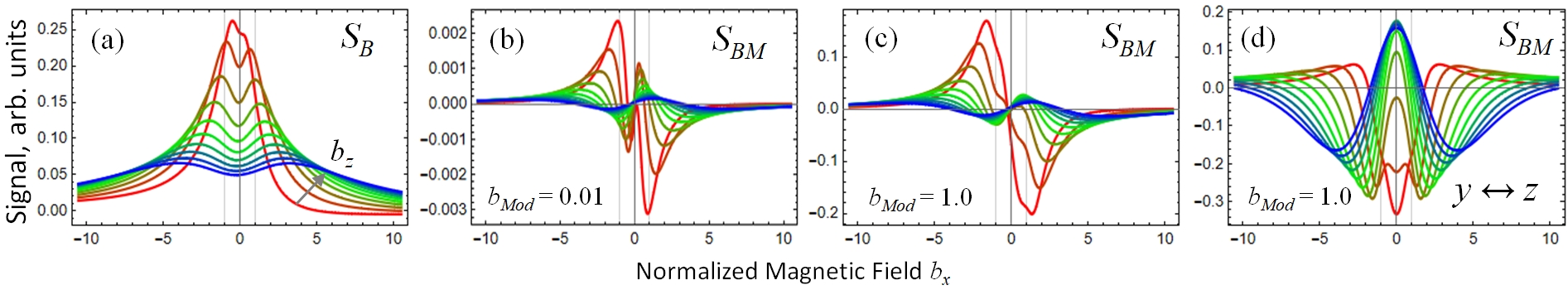}
	\caption{Calculated dependencies of the alignment balanced signals \cite{Meraki_Elson_Ho_Akbar_Kozbial_Kolodynski_Jensen_2023} on the normalized magnetic field bx ($\bf{x}||\bf{E}$, $\bf{k}||\bf{z}$) for the offset field values $b_y = 0.1, b_z  =  0.5, 1, 1.5,\ldots 5$ (from red to blue): (a) the balanced signals $S_B$; (b)–(d) signals $S_{BM}$, obtained with slow modulation of the field $b_x$ and demodulated using a lock-in amplifier. (b) normalized modulation amplitude $b_{Mod} = 0.01$,  (c) $b_{Mod} = 1.0$, (d) $b_{Mod} = 1.0$, the $y$ and $z$ axes swapped.}\label{figure1}
\end{figure*}

At the same time, it can be confidently stated that the potential of quantum alignment sensors has not yet been fully revealed. Thus, in  \cite{Meraki_Elson_Ho_Akbar_Kozbial_Kolodynski_Jensen_2023}, a cell with a paraffin coating was investigated. Such coatings retains their properties at relatively low (below 70~$^{\circ}$C) temperatures. Signals in such cells are characterized by extremely small widths  \cite{Graf_Kimball_Rochester_Kerner_Wong_Budker_Alexandrov_Balabas_Yashchuk_2005}, but also, as a consequence of low concentrations of atomic vapors, by relatively small amplitudes. In addition, cells with anti-relaxation coatings are very difficult to make compact. Therefore, in our work, signals were investigated in a compact (5$\times$5$\times$5~mm$^{3}$) cell with high buffer gas pressure – in such cells, the SERF effect in orientation is realized.

A rigorous definition of orientation, alignment, and higher-order moments is given within the framework of the so-called $\kappa$q-representation \cite{Blum_2012,Omont_1977}, based on the expansion of the momentum distribution in irreducible spherical operators \cite{Breschi_Weis_2012}:

\begin{equation}
    \rho = \sum_{\kappa=0}^{2F} \sum_{q=-\kappa}^{\kappa} \rho_q^{(\kappa)} T_q^\kappa
    \label{eq:equat_1}
\end{equation}

The convenience of the $\kappa$q-representation is that the dynamics of each of the multipoles into which the momentum distribution in the ensemble is decomposed can be described independently of the other multipoles. In particular, the second-order moment is characterized by five multipoles $\rho^{(2)}_i$  ($i = -2 \ldots 2$, the upper index corresponds to the alignment), each of which, in turn, can be characterized by its own relaxation rate. In the steady state, of these multipoles, only the multipole $\rho^{(2)}_{0} \sim$ $<3F_z^2-F^2>$, which is responsible for the population distribution \cite{Budker_Kimball_DeMille_2004} (p. 375), is non-zero – it is usually called the “alignment”. It is this multipole that is created and detected by linearly polarized pump light.

Let us consider a hydrogen-like atom with a nuclear spin $I$. The ground state of such an atom is split into two hyperfine levels with total moments $F = I \pm 1/2$, corresponding to two possible projections of the electron spin onto the nuclear spin. In a magnetic field, each hyperfine level is split into $2F + 1$ sublevels with projections of the magnetic moment $m_F = -F \ldots F$, characterized by relative populations $n_{mF}$. Then in a simple three-level system with $I = 1/2$ and $F = 1$, $m_{F} = -1, 0, +1$, and

\begin{equation}
    \rho_0^{(2)} = \frac{1}{\sqrt{6}}(n_{-1} - 2 \times n_0 + n_1).
    \label{eq:equat_2}
\end{equation}

Similar expressions can be derived for other values of $I$. It is easy to verify that in the equilibrium state $\rho^{(2)}_0 = 0$. The maximum value of $|\rho^{(2)}_0|$ is realized in two cases: $i$) all atoms are equally distributed over the levels $m_F= \pm F$ (${n_{-F}=n_{+F}=1/2}$)  and $ii$) all atoms are concentrated at the level $m_F = 0$ ($n_0 = 1$). The first case corresponds to the maximum positive alignment, the second to the maximum negative alignment. A theoretical study of the relaxation rates of moments of different multipolarity was carried out by W. Happer in \cite{Happer_1970}. It was shown, for example, that as a result of identical dipole perturbations, the quadrupole moment (i.e. alignment) relaxes three times faster than the dipole moment. Further in \cite{Breschi_Weis_2012,Meraki_Elson_Ho_Akbar_Kozbial_Kolodynski_Jensen_2023}, it was shown that at the same relaxation rate, the Hanle resonances in the alignment are characterized by widths that are half as large. But atoms were considered as an open system under external influence in both these works. On the contrary, in the context of this work, we are interested in relaxation in a quasi-closed system under the influence of internal factors (i.e. spin exchange).

It is clear that in such a system, collisions can destroy the alignment state without violating either the law of conservation of momentum or the selection rules. During collisions, atoms can exchange the directions of the electron spins. As a result of the exchange, the total moment of the system and its projection on the quantization axis are preserved, and the projection of the spin of each electron changes by $\pm 1$. Then, due to hyperfine interaction, the moment is redistributed between the electron $S$ and nuclear $I$ spins of each atom, and the state of the electron spin is randomized to a significant extent (since $I \ge S$) \cite{Happer_Tam_1977}.

Let us consider a system of two identical atoms with moment projections $m_{F1} = -F$ and $m_{F2} = F$ (maximum achievable positive alignment). The total moment of such a system, limited by the range $0...2F$, depends on the mutual directions of the two moments, and with the highest probability it is equal to zero (this also applies to an ensemble of atoms in a cell in the absence of circularly polarized light). The projection of the total moment onto the magnetic field is also equal to zero. As a result of a collision, the atoms can pass to the levels $m_{F1}  = -F + 1$ and $m_{F2}  = F - 1$, respectively; then, after randomization of the electron spin, collide again and either return to the previous state, or pass to the levels $m_{F1}  = -F + 2$ and $m_{F2}  = F - 2$, etc. In the case of initially negative alignment, atoms with moment projections ${m_{F1} = m_{F2} = 0}$ as a result of a collision can pass to the levels $m_{F1}$ = –1 and $m_{F2}$ = 1, respectively, etc. In all the cases considered, collisions in the absence of pumping will eventually lead to the restoration of the equilibrium population while maintaining zero angular momentum. The obvious conclusion is that spin-exchange relaxation of alignment in a closed system is possible even under the condition of conservation of the total momentum.

Thus, we have no a priori grounds to expect that in the case of alignment, spin-exchange relaxation will not contribute to time $T^{(2)}_{1}$. For the same reasons, we have no grounds to expect that in zero magnetic fields, the effect of transverse (time $T^{(2)}_{2}$) relaxation suppression similar to the SERF effect in orientation is realized in alignment.

However, our experimental results also suggest that \textit{i}) in ultra-weak, or ``zero'' magnetic fields (\textbar \textbf{B\textbar ~}\textit{$<$~$\Gamma $/$\gamma $}, where $\Gamma $ is the relaxation speed, and\textit{ }$\gamma $ is the gyromagnetic ratio), the alignment relaxes without any noticeable contribution from spin-exchange broadening, and \textit{ii}) in these fields, the alignment can exhibit resonances with a width approximately equal to the width of the ultra-narrow SERF resonances in the orientation. 

Orientation and alignment are multipole moments of the angular momentum distribution. In a closed system, the conservation of orientation (i.e., the angular momentum) follows from the isotropy of space, i.e., invariance under rotations (Noether's theorem). Conservation of alignment, i.e., the quadrupole moment, requires a symmetry of a higher order than the standard space-time symmetries, so alignment is not considered a conserved quantity. Nevertheless, the results of this work may suggest the existence of fundamental reasons for the conservation of alignment, similar to the reasons for the conservation of orientation.

The remainder of the paper is organized as follows: Section \ref{sec:2} provides a brief theoretical background, Section \ref{sec:3} describes the experimental setup, Section \ref{sec:4} presents the experimental results, and Section \ref{sec:5} discusses them.

\section{Theory}\label{sec:2}

The theory describing Hanle resonances in alignment was developed in detail in \cite{Breschi_Weis_2012} and subsequently in \cite{Bevilacqua_Breschi_Weis_2014,LeGal_Palacios-Laloy_2022,Meraki_Elson_Ho_Akbar_Kozbial_Kolodynski_Jensen_2023}. According to \cite{Breschi_Weis_2012}, in the $xyz$ coordinate system related to the direction of beam polarization ($\bf{E}||\bf{x}$, $\bf{k}||\bf{z}$, where $\bf{k}$ is wave vector)

\begin{equation}
    \frac{\rho_{0}^{(2)}}{\rho_{0 Eq}^{(2)}} = \frac{1}{4} + \frac{3}{4} \frac{16 b_{x}^{4} + 8 b_{x}^{2} + 1}{4 b_{x}^{4} + 4 b_{yz}^{2} + 1} - 3 \frac{b_{x}^{4} + b_{x}^{2}}{b_{x}^{4} + b_{yz}^{2} + 1},
    \label{eq:equat_3}
\end{equation}
where $\rho^{(2)}_{0 Eq}$ is the equilibrium value of the alignment multipole, $b_x = B_x/B_0$, $b_{yz} = \sqrt{(B_y^2 + B_z^2)}/B_0$, $B_0 = \Gamma/\gamma$.


In a strong magnetic field ($|\bf{B}| \gg \Gamma/\gamma$) the precessing alignment multipoles $\rho_i^{(2)}$ ($i$ = –2, –1, 1, 2) do not contribute to the DC signal. In this case the main manifestation of alignment is linear dichroism due to $\rho_0^{(2)}$. It can be detected by measuring either the total intensity of the transmitted light or the rotation of its polarization due to anisotropic absorption. Hereinafter we will call the corresponding signals $S_T$ (total) and $S_B$ (balanced) \cite{Akbar_Kozbial_Elson_Meraki_Kolodynski_Jensen_2024,Meraki_Elson_Ho_Akbar_Kozbial_Kolodynski_Jensen_2023}. We will be interested only in balanced signals, since they correspond to a significantly higher signal-to-noise ratio. 

As shown in Appendix~A, in the standard experimental geometry $\bf{B}||\bf{x}$ or $\bf{B}||\bf{y}$ the $S_B$ signal due to linear dichroism is zero (this also follows directly from the symmetry of the system). However, it will be shown below (see Figure~\ref{figure4}) that when recording Hanle resonances the maximum amplitude of the $S_B$ signal is achieved at $\bf{B}||\bf{x}$, or $\varphi = \arctan (b_{y}/b_x) = 0$. An explanation is given in \cite{Meraki_Elson_Ho_Akbar_Kozbial_Kolodynski_Jensen_2023}: for $|\bf{B}| \gg \Gamma/\gamma$, the main contribution to the stationary rotation of polarization is given by the non-diagonal components of the alignment tensor:

\begin{equation}
    S_{B} \sim i(\rho _{1}^{(2)} +\rho _{-1}^{(2)})
    \label{eq:equat_4}
\end{equation}
while the main contribution to the absorption signal  $S_T$ (studied in \cite{Breschi_Weis_2012}) is still made by the multipole $\rho^2_{0}$. 

The physical basis for this is as follows: under some conditions (e.g. at $\varphi = 0, \pm\pi/2$) the $S_B$ signal is insensitive to changes in the alignment multipole $\rho ^2_{0}$, and the Hanle signals are formed entirely by precessing multipoles, the precession of which in the ``zero'' field turns into a stationary deviation.  The angle of deviation is determined by the ratio of the precession rate to the relaxation rate. A formula describing the balanced signal is given in \cite{Meraki_Elson_Ho_Akbar_Kozbial_Kolodynski_Jensen_2023}. After translation into the coordinate system we use, it takes the form

\begin{multline}
   S_{B} \sim \\ \sim \frac{b_{z} \left(1+4b_{x}^{4} +b_{y}^{2} +b_{z}^{2} \right)-b_{x} b_{y} \left(1+2b_{x}^{4} -2b_{y}^{2} -2b_{z}^{2} \right)}{\left(4b_{x}^{4} +4b_{y}^{2} +4b_{z}^{2} +1\right)\left(b_{x}^{4} +b_{y}^{2} +b_{z}^{2} +1\right)}
    \label{eq:equat_5}.
  \end{multline}

From Eq.~\eqref{eq:equat_5} it follows that at $b_{i} \ll 1$ $({i = x, y, z})$ the $S_{B}$ signal vanishes only when the condition $b_{z}=b_{x} \cdot b_{y}$ is met, but not when any one component of the field is zeroed.

The signals calculated according to \cite{Meraki_Elson_Ho_Akbar_Kozbial_Kolodynski_Jensen_2023} for the case when the \textit{b${}_{x}$} field is scanned and the offset field is applied along the \textit{z} axis or the \textit{y} axis are shown in Figure~\ref{figure1}. As we can see, in the latter case (Figure~\ref{figure1}d) the signals change their shape from antisymmetric to symmetric. It is easy to show that this does not happen with signals calculated according to Eq.~\eqref{eq:equat_3}, so the response of signals to the rotation of the field in the $0xy$ plane allows us to distinguish signals due to the multipole $\rho^{2}_{0}$ (i.e., linear dichroism) from signals due to the off-diagonal components of the alignment tensor.

In all the above expressions, the dependence of the value of $\Gamma$ on external factors was not taken into account. Such dependencies for orientation were derived in \cite{Appelt_Ben-AmarBaranga_Young_Happer_1999}. In particular, it was shown that in vicinity of ``zero'' fields ($|\textbf{B}| \ll \Gamma /\gamma$) the widths $\Delta$ and frequencies $\omega$ of magnetic resonance are determined by the expressions

\begin{equation}
   \Delta =\mbox{Re}\left(\omega \right),\omega =\mbox{Im}\left(\omega \right),
    \label{eq:equat_6}
\end{equation}

\begin{multline}
  \omega =\frac{\left[I\right]^{2} +2}{3\left[I\right]^{2} } R_{SE} -\\
  -\sqrt{-\omega _{0}^{2} -\frac{2i\omega _{0} R_{SE} }{\left[I\right]} +\left(\frac{\left(\left[I\right]^{2} +2\right)R_{SE} }{3\left[I\right]^{2} } \right)^{2} },
\label{eq:equat_7}
\end{multline}
where $\left[I\right]=2 I+1$, $\omega_{0} =g_{s} \mu _{B} B/(\hbar \left[I\right])$, \textit{I} is the nuclear spin, \textit{R${}_{SE}$} is the spin-exchange rate, \textit{g${}_{s}$} is g-factor, \textit{$\mu $${}_{B}$} is the Bohr magneton (quoted from \cite{Seltzer_2008}, p.141). As stated in the Introduction, we have no apriori reason to believe that the dependence of alignment on the magnetic field will obey the same (or similar) laws, but, as we will see later, the comparison can be quite informative.

\begin{figure*}[!t]  
	\includegraphics[width=\linewidth]{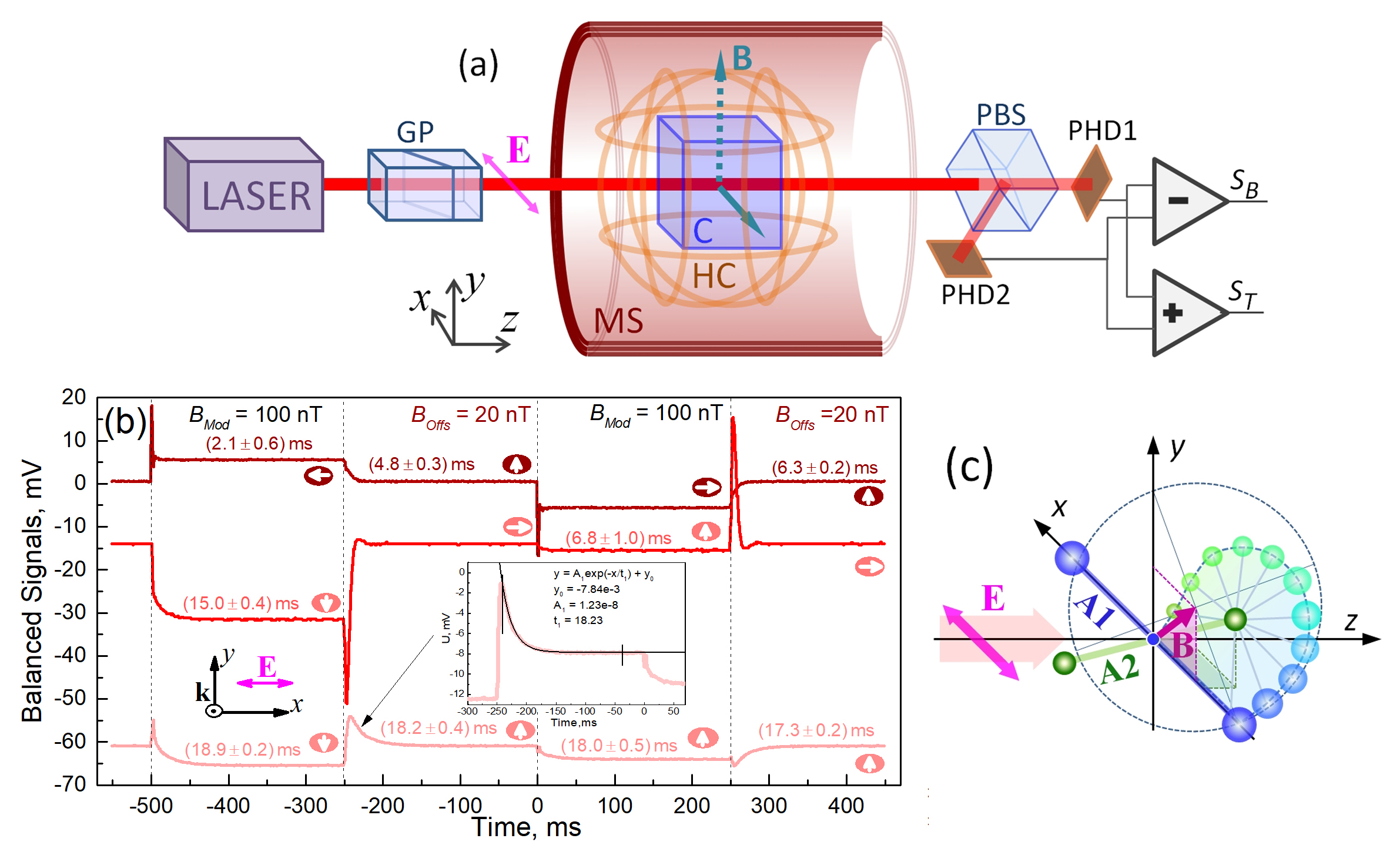}
	\caption{(a) Experimental setup: MS -- magnetic shield, GP -- Glan prism, HC -- Helmholtz coils system, C -- cell, PBS -- polarization cube, PHD1, PHD2 -- photodiodes. The laser frequency stabilization circuit is not shown. (b) Examples of recordings of polarization rotation signals when switching the induction and direction of the magnetic field. The vertical shift of the graphs is arbitrary. Arrows designate the direction of the magnetic field in the $0xy$ plane (polarization vector $\textbf{E}||\textbf{x}$, wave vector $\textbf{k} || \textbf{z}$) in a given recording interval, the numbers above the arrows are the modulus of the magnetic field induction. Several values of the relaxation times are given. More recordings are provided in Appendix B. The inset shows an example of fitting a transient process with an exponential dependence. (c) Dynamics of alignment when turning on the ultra-weak field \textbf{B} in the $0xy$ plane in case of $T^{(2)}_1  \approx  T^{(2)}_2 \approx 1/\omega_{L}$. A1 is the initial state of alignment along the $x$ axis; ten intermediate states are also shown. A2 is the state of alignment in the stationary case at constant pumping and magnetic field; it is obtained by averaging over the intermediate states. The dashed arcs are the hodograph of the precession of alignment in the absence of relaxation.}\label{figure2} 
\end{figure*}

\section{Experiment}\label{sec:3} 

The experimental setup, previously described in \cite{Petrenko_Pazgalev_Vershovskii_2024}, is shown in Figure~\ref{figure2}a. Below we use a coordinate system in which the beam propagates along the \textit{z} axis, and the electric component \textbf{E} of the light is parallel to the \textit{x} axis.

The cell chosen for the study contained, in addition to Cs, $P_{N2} \approx 300$~torr of buffer gas (nitrogen), which ensured complete mixing of the excited states of Cs. The external cavity diode laser (VitaWave company, Moscow) was tuned to the center of the 
$F=3 \to F'=3$ transition of \textit{D}${}_{1}$ absorption line of Cs (wavelength 895 nm). A Glan prism was used to improve the degree of beam polarization. 

The cell was placed in an oven (not shown in the scheme), which in turn was placed in the center of a multilayer magnetic shield. The residual magnetic field was compensated by the coils of the three-coordinate Helmholtz system using the criterion of the absence of the beam polarization rotation, i.e. up to the light shift of the magnetic levels in the system. The alignment signals were detected by a balanced photodetector, which allowed both the total (\textit{S${}_{T}$}) and the differential, or balanced (\textit{S${}_{B}$}) signals to be simultaneously recorded.

The easiest way to measure \textit{S${}_{B}$} is to split the beam into two perpendicularly polarized components with initially equal intensities. The axis of the polarization-separating element is set at an angle of $\pi $/4 to the initial azimuth of the beam polarization, the intensity of the components is measured by two photodetectors, the signals of which are subtracted to measure the \textit{S${}_{B}$} signal, and can be summed to measure the \textit{S${}_{T}$} signal. Using balanced signals allows the signal-to-noise ratio to be significantly increased due to subtraction of technical laser noise. The total and the differential signals from the photodiodes are amplified by transimpedance amplifiers. 

The coefficient of conversion of the photocurrent into the photodetector output voltage  (shown below in the graphs) was 5$\times$10${}^{3}$~V/A.

The results shown in Figure~\ref{figure2}b were obtained at a cell temperature of \textit{T${}_{c}$} = 120${}^{o}$C. The calculated parameters at this temperature are as follows: the atomic concentration is 4.75$\times$10${}^{13}$~cm${}^{-3}$ \cite{Margrave_1964}, the spin-exchange relaxation rate is 47$\times$10${}^{3}$~s${}^{-1}$ \cite{Vershovskii_Dmitriev_Petrenko_2021}.

For comparison with the results of our experiment presented below, we note that the typical linewidth of the ``classical'' (i.e. non-zero field) magnetic resonance in orientation, calculated taking into account the nuclear slowing factor, is approximately 1.6~kHz. This width, converted to the magnetic field scale using the Cs gyromagnetic ratio $\gamma $~=~3.5~Hz/nT, is 460~nT \cite{Seltzer_2008}. The corresponding time constant is \~{}0.1~ms.

\section{Results}\label{sec:4} 

Figure \ref{figure2}b shows an example of the records of \textit{S${}_{B}$} signals registered when switching the strong transverse field $\bf{B}_{\tt{Mod}}$ (${B}_{Mod}$~=~100~nT) on and off in the presence of the weak bias field $\bf{B}_{\tt{Offs}}$ (${B}_{Offs}$~=~20~nT); both fields lie in the 0\textit{xy} plane. In other words, the magnetic field was switched from $\bf{B}_{\tt{Offs}}$ to $\bf{B}_{\tt{Offs}}$~+~$\bf{B}_{\tt{Mod}}$  ($B_{Mod} \gg B_{Offs}$). At this stage, lock-in amplification was not applied, and signals were taken directly from the output of the transimpedance amplifier. Additional records are provided in Appendix B. These data are presented mainly to show that different configurations of switching fields correspond to different kinds of transient effects, as well as to the sections with slow (up to 18~ms) relaxation. This relaxation time is about two orders of magnitude longer than the spin-exchange relaxation time. It roughly corresponds to the values of $T^{(1)}_{1}$ calculated for \textit{orientation} signals according to \cite{Seltzer_2008} for the conditions listed in the Experiment section. The results of calculating the contributions of specific processes (spin-destructive collisions, collisions with buffer gas, wall collisions) for a wide range of temperatures are shown below in Figure~\ref{figure7}. 

There would be nothing surprising in this if it were not for the fact that the absence of a contribution from spin exchange to the longitudinal relaxation rate $T^{(1)}_{1}$ is due solely to the law of conservation of angular momentum \cite{Seltzer_2008}. However, as has already been noted, the destruction of the alignment multipoles $\rho^{(2)}_{i}$ is not associated with a change in average angular momentum; therefore we have no grounds to expect that in the case of alignment, spin-exchange relaxation will not contribute to time $T^{(2)}_{1}$. Moreover, it would be natural to expect that strong spin exchange would quickly destroy the multipoles $\rho^{(2)}_{i}$.

Further studies were carried out in ``zero'' field range. The experimental setup remained the same: we superimposed a switching field \textbf{B${}_{\tt{Mod}}$} on a DC bias field \textbf{B${}_{\tt{Offs}}$}, but now \textit{B${}_{Mod}$}~was fixed at 5~nT, and \textit{B${}_{Offs}$}~varied between 0 and 50~nT (Figure~\ref{figure3}a). More examples are provided in Appendix B.

In addition to switching the field \textbf{B${}_{\tt{Mod}}$} direction to the opposite ($x \Leftrightarrow x$ and $y \Leftrightarrow y$), we changed the direction of the \textbf{B${}_{\tt{Mod}}$} field in the $0xy$ plane by $\pi $/2 ($x \Leftrightarrow y$) (as in \cite{Petrenko_Pazgalev_Vershovskii_2024}). Figure~\ref{figure3}b shows the relaxation times measured after the end of the transient (precession) period. The projection of the total field \textbf{B${}_{\tt{Offs}}$}~+~\textbf{B${}_{\tt{Mod}}$} onto the bias field \textbf{B${}_{\tt{Offs}}$} is plotted along the horizontal axis. 

\begin{figure}[!t]  
	\includegraphics[width=\linewidth]{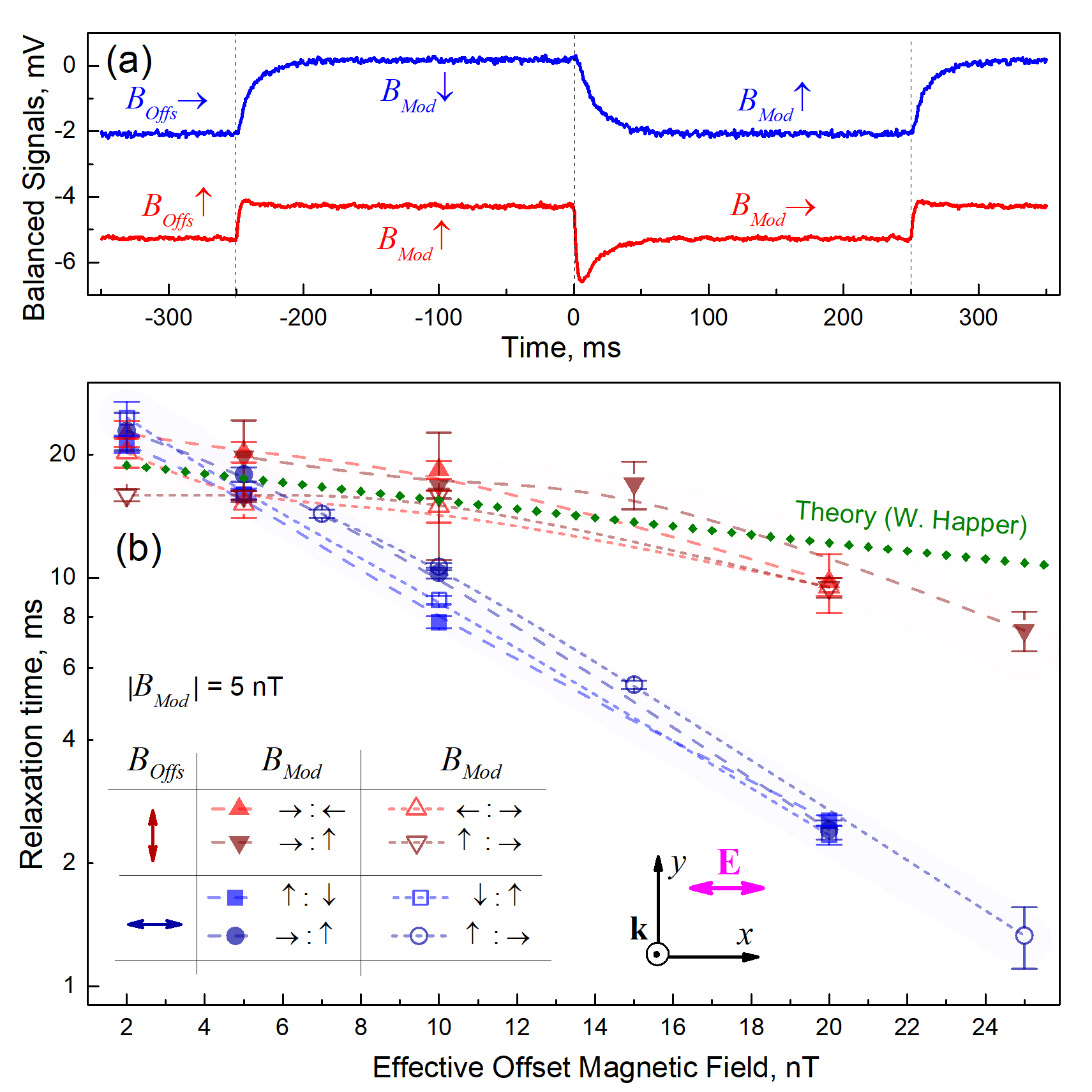}
	\caption{(a) Examples of signal recordings made during magnetic field $\bf{B}_{\tt{Mod}}$ switching (${B}_{Mod}$~=~5~nT). Arrows indicate the direction of the switching  in the 0\textit{xy} plane modulation magnetic field $\bf{B}_{\tt{Mod}}$ and the DC bias field $\bf{B}_{\tt{Offs}}$ (${B}_{Offs}$~=~5~nT) in a given recording interval; the direction of \textbf{E} is considered horizontal (\textbf{E}\textbar \textbar \textbf{x}, \textbf{k}\textbar \textbar \textbf{z}). More recordings are provided in Appendix B (Figure~\ref{figureB2})). (b)~Relaxation times measured after the end of the precession process and plotted against the projection of the total field onto the direction of the bias field $\bf{B}_{\tt{Offs}}$. Dark green diamonds are the theoretical dependence of the SERF orientation resonance widths, calculated according \cite{Seltzer_2008} (p.141). The raw data are provided in Appendix B (Figure~\ref{figureB2}).}
    \label{figure3}
\end{figure}

\begin{figure}[!t]  
	\includegraphics[width=\linewidth]{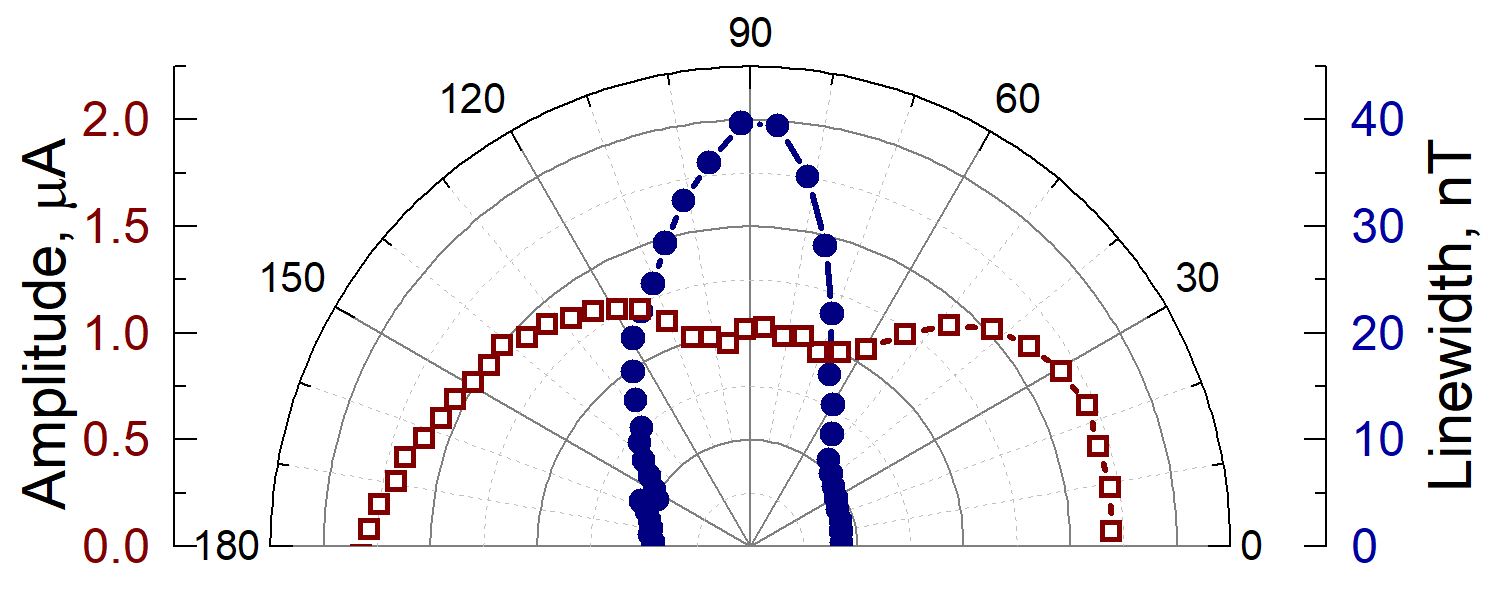}
	\caption{Parameters of alignment signals at \textit{T~}=~107${}^{o}$C under modulation and scanning of the \textit{B} field in the 0\textit{xy} plane depending on the azimuth angle \textit{$\varphi$}: open squares are the resonance amplitudes, closed circles are the resonance widths (HWHM).}
    \label{figure4}
\end{figure} 

The scheme with switching the direction of the small modulating field \textbf{B${}_{\tt{Mod}}$} in a ``zero'' constant field is essentially identical to the scheme of the Hanle magnetometer, according to which SERF sensors are built \cite{Kominis_Kornack_Allred_Romalis_2003,Seltzer_Romalis_2004}, with the exception that in our experiment we use linearly polarized light (as in \cite{Meraki_Elson_Ho_Akbar_Kozbial_Kolodynski_Jensen_2023,LeGal_Palacios-Laloy_2022}). However, in a real magnetometer scheme, the lock-in amplification method is usually used to register the signal. It allows the signal to be transferred to the modulation frequency, thereby increasing the signal-to-noise ratio. The demodulated signals arising when scanning one magnetic field component around zero value have a resonant shape (Hanle resonances). They are characterized by a width \textit{$\Delta $${}^{(i)}$} (\textit{i}~=~1,2) determined by the inverse time $T^{(1)}_{2}$ for \textit{orientation}, and, correspondingly, $T^{(2)}_{2}$ for \textit{alignment} ($\Delta^{(1)} \sim 1/T^{(1)}_{2}$, $\Delta^{(2)} \sim 1/(2T^{(2)}_{2})$) \cite{Breschi_Weis_2012, Meraki_Elson_Ho_Akbar_Kozbial_Kolodynski_Jensen_2023, Seltzer_2008}. 

The experiments described below were performed using standard lock-in techniques. Sinusoidal modulation of the fields in the \textit{X} and/or \textit{Y} coils with a frequency of 5~Hz and an amplitude of 2.5~nT (5~nT p-t-p) was used. The \textit{S${}_{B}$} signals were extracted by a lock-in amplifier at the field modulation frequency. Next, the obtained signals were approximated by Eq.~\eqref{eq:equat_5}, and/or by a sum of Lorentz and dispersion contours, with coefficients cos(\textit{$\psi $}) and sin(\textit{$\psi $}) corresponding to the phase shift \textit{$\psi $} in the system. 

We investigated the parameters of Hanle resonances in alignment at different cell temperatures, magnetic field values and modulation parameters. First, we measured their angular dependences (Figure \ref{figure4}); some asymmetry in the distribution is due to the presence of residual fields. Note that the resonance width depends very sharply on the azimuthal angle, and it is minimal at $\varphi $~=~0, i.e., when scanning along the beam polarization direction (\textbf{E}\textbar \textbar \textbf{x}). 

Therefore, we proceeded with recording the resonances by linearly scanning \textbf{B${}_{x}$} at a fixed value of the bias field \textbf{B${}_{z}$}. We varied the bias field to see an effect similar to the SERF effect in orientation. The choice of the \textit{z} direction for the bias field provided an additional guarantee that we were observing alignment signals rather than orientation signals, since the Hanle resonances in orientation are almost insensitive to the \textit{z}-axis field (this field destroys the Hanle mode but does not cause a resonant response). A small field along the \textit{y}-axis ($B_{y} \approx 3$~nT) was applied to ensure sensitivity to alignment signals at $B_{z} \approx 0$. 

\begin{figure}[!t]  
	\includegraphics[width=\linewidth]{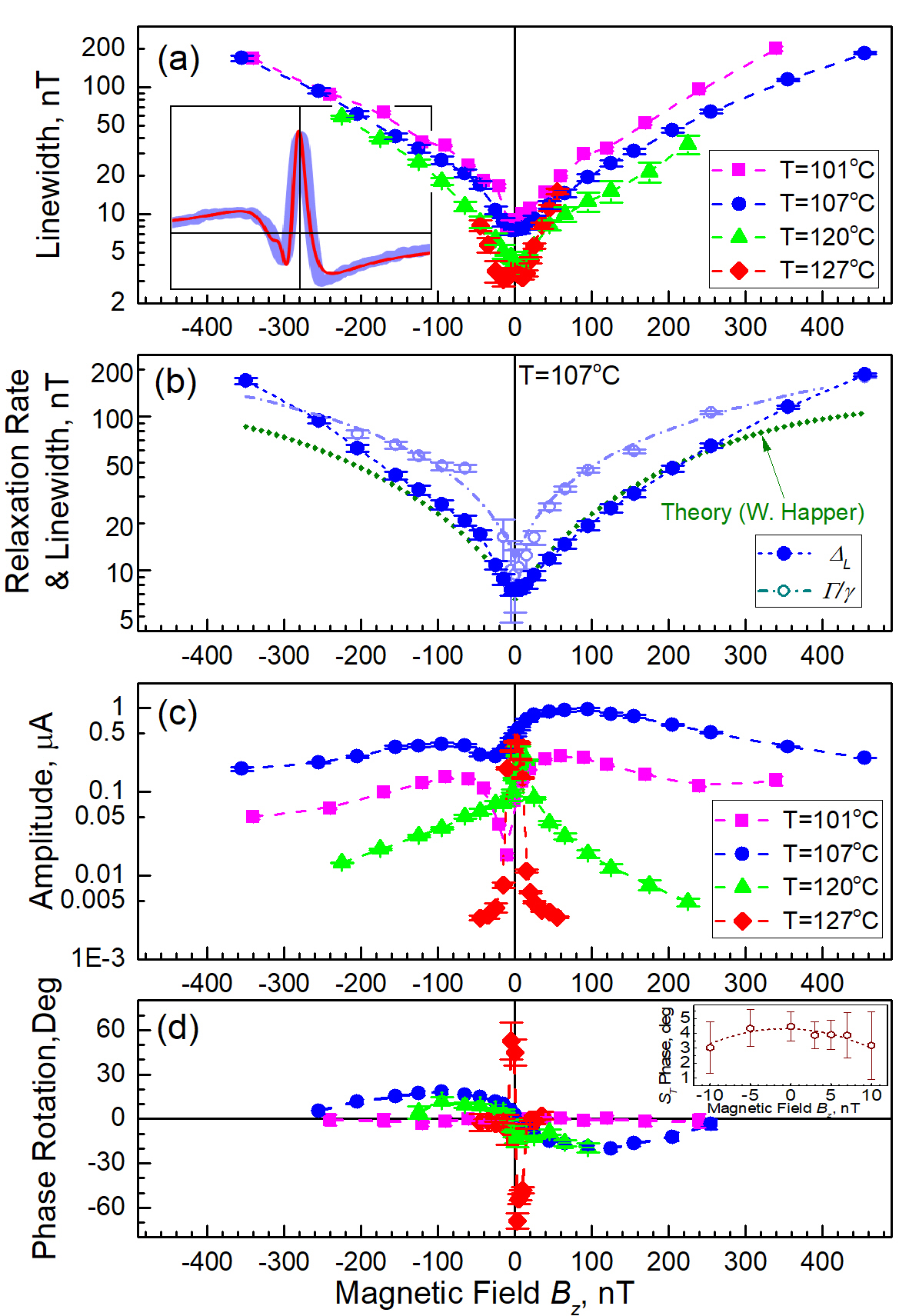}
	\caption{Parameters of alignment signals under modulation and scanning of the \textit{B${}_{x}$} field depending on the values of the field \textit{B${}_{z}$}: (a) the resonance widths (HWHM) $\Delta_{L}$ obtained by fitting with Lorentz contours. The inset shows an example of a record (light blue) with approximations using Eq. \eqref{eq:equat_5} \cite{Meraki_Elson_Ho_Akbar_Kozbial_Kolodynski_Jensen_2023} (red); (b) the widths $\Delta_{L}$ and the rates $\Gamma /\gamma$ obtained by fitting signals with Eq.~\eqref{eq:equat_5} at \textit{T}~=~107~${}^{o}$C; dark green diamonds are the theoretical dependence of the SERF orientation resonance widths, constructed according to Eqs. \eqref{eq:equat_6}, \eqref{eq:equat_7} \cite{Seltzer_2008}; (c) the amplitudes of \textit{S${}_{B}$} signal; (d) rotation of the phase of \textit{S${}_{B}$} signal; the inset -- rotation of the phase of \textit{S${}_{T}$} signal at 127${}^{o}$C. The lines are guides for the eyes. The raw data are provided in Appendix B (Figures \ref{figureB3}, \ref{figureB4}).}
    \label{figure5}
\end{figure}

Figure~\ref{figure5} shows the results of the study of the near-zero-field resonance parameters at different cell temperatures. As noted above, the shape of Hanle resonances at high temperatures can differ significantly from the classical one. In our experiment we observed all the effects mentioned in \cite{DiDomenico_Bison_Groeger_Knowles_Pazgalev_Rebetez_Saudan_Weis_2006, Meraki_Elson_Ho_Akbar_Kozbial_Kolodynski_Jensen_2023, Breschi_Weis_2012}. For the demonstration in Figure~\ref{figure5}, conditions (namely, modulation along \textit{x}-axis) were chosen in which the differences in the resonance shape from the Lorentzian were not too significant.  

\begin{figure}[!t]  
	\includegraphics[width=\linewidth]{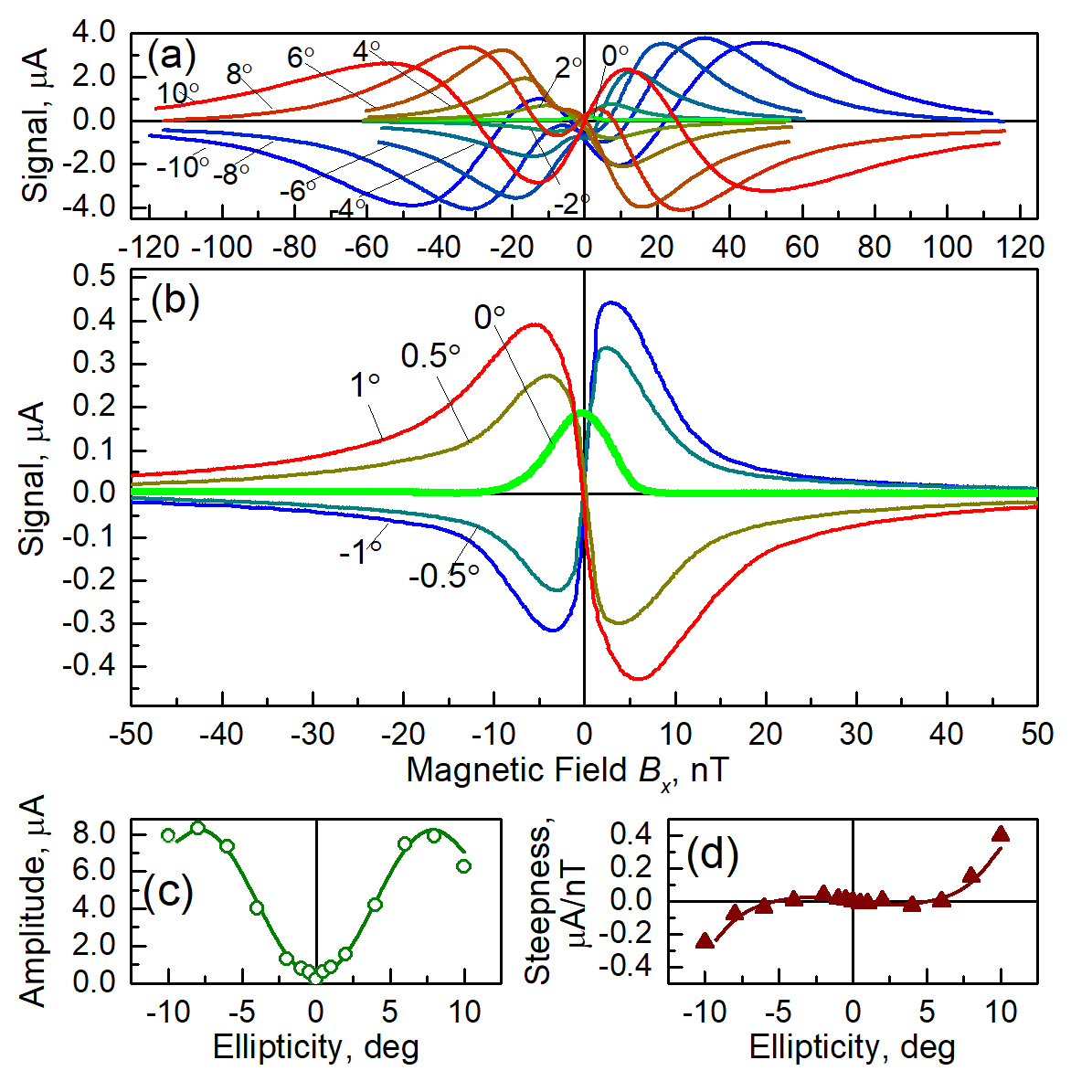}
	\caption{Recordings of demodulated \textit{S${}_{B}$} signals depending on the \textit{B${}_{x}$} field at different degrees of laser light ellipticity: (a) ellipticity varies within the range of -10 -- +10${}^{o}$, (b) -- within the range of -1 -- +1${}^{o}$, (c) the signal amplitudes, (d) their steepness at \textit{B${}_{x}$} = 0; the lines in (c),(d) are polynomial approximations.}\label{figure6}
\end{figure}

To ensure that the narrow resonances recorded in our experiment in ``zero'' fields were indeed alignment signals, we conducted a control experiment: using the additional quarter-wave plate, we introduced into the light ellipticity controlled by a Thorlabs polarimeter. The results are shown in Figure~\ref{figure6}.

\section{Discussion}\label{sec:5} 

To interpret the obtained results, we turn to the features of the formation of the alignment signal. The hodograph of the collective atomic moment in the case of pure alignment forms a symmetrical figure -- either, depending on the alignment sign, a ``barbell'' or a ``donut'' \cite{Budker_Kimball_DeMille_2004, Rochester_Ledbetter_Zigdon_Wilson-Gordon_Budker_2012}. Figure~\ref{figure2}c schematically shows the dynamics of the alignment in the system with approximately equal relaxation times $T^{(2)}_{1} \approx T^{(2)}_{2} \approx 1/\omega_{L}$ when the field \textbf{B} in the $0xy$ plane is switched on. 
Initially (in the absence of the field), the light creates alignment (let it be a ``barbell'') in the direction of the vector of the electric component \textbf{E\textbar \textbar x}. As in the case of the ``classical'' Hanle effect in orientation, the result of the combined action of pumping, precession and relaxation is a turn of the ensemble, resulting in the appearance of non-zero alignment components along the \textit{y} and \textit{z} axes, and in a decrease in alignment along the \textit{x} axis. The latter effect is detected by the pump beam. In this case, the transient processes caused by precession can significantly enhance the magnitude of the observed signals, which is observed in Figure~\ref{figure2}b at sufficiently large fields. Apparently, this conclusion corresponds to the theoretical description \cite{Meraki_Elson_Ho_Akbar_Kozbial_Kolodynski_Jensen_2023}.

Note that the records in Figure~\ref{figure2}b, made with an $x \Leftrightarrow y$ change in the field direction, are also characterized by a fast (compared to the times $T^{(2)}_{1}$, $T^{(2)}_{2}$, $1/\omega_{L}$) jump associated with a change in the detecting properties of the light. An analysis of the records presented in Figure \ref{figure3} and Figure \ref{figureB2} shows that for \textit{B${}_{Offs}$~$>$~B${}_{Mod}$} they can be divided into two groups corresponding to two directions of the \textbf{B${}_{\tt{Offs}}$} field. The greatest degree of similarity of the records in the two groups is achieved if the effective offset field, i.e. the projection of the total field \textbf{B}~=\textbf{ B${}_{\tt{Offs}}$}~+~\textbf{B${}_{\tt{Mod}}$} onto the \textbf{B${}_{\tt{Offs}}$} direction (Figure \ref{figure3}b), is plotted on the horizontal axis. For comparison, Figure \ref{figure3}b shows the theoretical dependence of the SERF \textit{orientation} resonance width \cite{Happer_Tam_1977}, constructed according to \cite{Seltzer_2008}  (p.141) under the following conditions: \textit{T${}_{c}$}~= 120${}^{o}$C, \textit{P${}_{N2}$}~= 300~torr, \textit{I${}_{p}$}~=~5~mW/cm${}^{2}$. It is evident that it qualitatively and quantitatively corresponds to the results we obtained for \textbf{B${}_{\tt{Offs}}$}\textbar \textbar \textbf{y}. 

For \textbf{B${}_{\tt{Offs}}$}\textbar \textbar \textbf{x}, the dependence of the relaxation time on the bias field is much sharper. It is well described by the exponential $T = T_{0} \cdot e^{-B/B_{0}}$, where \textit{T${}_{0}$}~=~32~ms, \textit{B${}_{0}$}~=~7.8~nT. Consequently, additional mechanisms are responsible for the destruction of the alignment under these conditions, and the rate of relaxation caused by them in ultra-weak fields depends exponentially (and not quadratically, as in the usual SERF) on the field strength. 

\begin{figure}[!t]  
	\includegraphics[width=\linewidth]{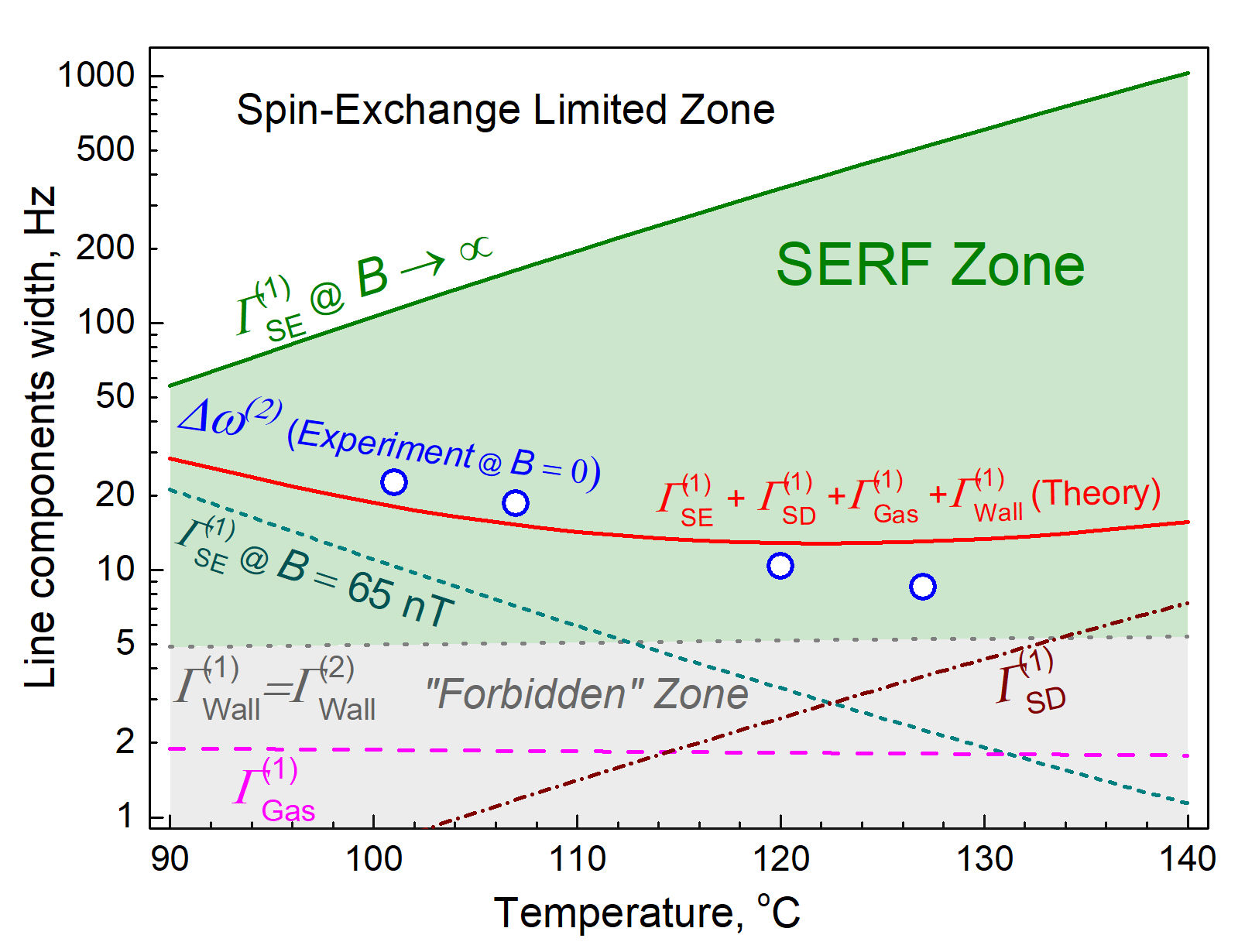}
	\caption{Calculated according to Seltzer\_2008 relaxation rates of orientation signals (lines) and their comparison with experimental data: \textit{$\Gamma $${}_{SE}$} is the spin exchange contribution, \textit{$\Gamma $${}_{SD}$} is the contribution of spin-destroying interactions, \textit{$\Gamma $${}_{Gas}$} is the contribution of collisions with the buffer gas, \textit{$\Gamma $${}_{Wall}$} is the contribution of wall collisions (the same for orientation and alignment). Upper indexes: Upper indexes: $(1)$ -– orientation, $(2)$ –- alignment. Calculation parameters: length of the edge of the cubic cell is 5~mm, nitrogen pressure is 300~Torr, degree of polarization of the medium is 0.1. The area shaded in gray is the area of forbidden width values, the area shaded in green is the SERF area. The blue circles are the resonance linewidths measured at \textit{B${}_{z}$}~=~0.}\label{figure7}
\end{figure} 

The maximum relaxation times recorded in our experiment are about (25~$\pm$~2)~ms, which should correspond to a resonance linewidth HWHM of about 6~Hz or 1.7--4.7~nT, depending on the atomic polarization degree \cite{Seltzer_2008} (in case of alignment -- twice as little \cite{LeGal_Palacios-Laloy_2022}). Recall that this is almost two orders of magnitude smaller than the linewidth of the ``classical'' magnetic resonance (approximately 1.6~kHz \cite{Seltzer_2008}) calculated without taking into account the SERF effect. On the other hand, as follows from Figure~\ref{figure3}b, the times obtained in the experiment are in good agreement with the orientation resonance widths calculated taking into account the SERF effect. It should be noted that in a zero field we observed very small light broadening effects (at the level of 2~nT/mW).

By analyzing angular dependencies (Figure~\ref{figure4}), we can make two main conclusions: first, the angular dependence of the amplitude is quite smooth, and the signal does not vanish at \textit{$\varphi $}~=~0, $\pm$$\pi $/2. Therefore, it definitely cannot be described by Eqs. \eqref{eq:equat_3}, \eqref{eq:equat_4} \cite{Breschi_Weis_2012}. Second, the resonance width depends very sharply on the azimuthal angle, and it is minimal at $\varphi $~=~0, i.e., when scanning along the beam polarization direction (\textbf{E}\textbar \textbar \textbf{x}). Note that in this configuration  the bias field \textbf{B${}_{\tt{Offs}}$} is directed along the \textit{y}-axis, which is consistent with the data in Figure~\ref{figure3}b.

The next graph (Figure~\ref{figure5}) demonstrates that large relaxation times do correspond to ultra-narrow resonances (Figure~\ref{figure5}a) that can be used in magnetometry. 

Figure~\ref{figure5}b shows an example of the comparison of the resonance widths (HWHM) $\Delta_{L}$ obtained by approximation with Lorentz contours with the relaxation rates \textit{$\Gamma $} obtained by approximation using Eq.~\eqref{eq:equat_5} according to \cite{Meraki_Elson_Ho_Akbar_Kozbial_Kolodynski_Jensen_2023}, taking into account the broadening by the \textit{b${}_{xy}$} field and the modulation amplitude (\textit{B${}_{Mod}$}~=~2.5~nT). Note that the approximation using Eq.~\eqref{eq:equat_5} is much more complex and requires careful adjustment of the initial values. The data in Figure~\ref{figure5}b confirm that with decreasing magnetic field, not only the apparent resonance width (as in Figure~\ref{figure1}a) but also the relaxation rate itself decreases. 

The dependence shown in Figure~\ref{figure5}b is superimposed on the theoretical dependence of the width of the SERF orientation resonances, constructed according to the Happer's theory (Eqs.\eqref{eq:equat_6},\eqref{eq:equat_7}, cited from \cite{Seltzer_2008} p.141). It is evident that qualitatively (up to the uncertainty of the effective temperature in the cell, light broadening, gas pressure, etc.) it describes our results very well (note that Eqs.\eqref{eq:equat_6},\eqref{eq:equat_7} describe not the relaxation rate, but rather the apparent resonance width $\Delta_{L}$).

It is worth noting the sharp decrease in the width of the amplitude distribution peak (Figure~\ref{figure5}c) at T~=~127${}^{o}$C -- the half-width of its distribution is about 7~nT. Some asymmetry in the results Figure~\ref{figure5}c is possibly due to \textit{i}) phase shift in the process of lock-in amplification, -- especially in the vicinity of zero field, where the linewidth becomes comparable with the frequency and amplitude of modulation, and \textit{ii})  presence of residual field along \textit{y}-axis (\~{}3~nT).\textit{ }

The dependencies of the signal phase on \textit{B${}_{z}$} in the vicinity of zero field have a dispersion form with a change in sign at $B_{z} \approx 0$ (Figure~\ref{figure5}d). At a temperature of 127${}^{o}$C an anomalous large change in the signal phase is observed in the vicinity of zero \textit{B${}_{z}$} field. It is associated with the presence of a small offset transverse field $B_{y} \approx 3$~nT, comparable in magnitude to the resonance width. The very change in the signal shape from antisymmetric to symmetric at \textit{B${}_{z}$} $<$ \textit{B${}_{y}$} (see Figures \ref{figure1}, \ref{figureB3}, \ref{figureB4}) clearly confirms that the signals we observe are described by Eq.  \eqref{eq:equat_5}, and therefore are coherence signals due to $\rho^{(2)}_{\pm1}$. Note that the phase of the \textit{S${}_{T}$} absorption signal remains virtually unchanged (see the inset in Figure~\ref{figure5}d). Signals recorded at lower temperatures do not show any peculiarities, therefore, the narrowing of the line in the vicinity of zero is not necessarily accompanied by a sharp change in phase. 

The most significant result following from Figure~\ref{figure5} is the decrease in the resonance width in zero field with increasing temperature, which is characteristic of the SERF effect.

The question may arise whether the signals recorded in our experiment are really alignment signals. Thus, in \cite{Rochester_Ledbetter_Zigdon_Wilson-Gordon_Budker_2012} it is shown that under certain conditions, e.g. with a rapid change in the electric field, dynamic effects lead to the of alignment-to-orientation conversion (AOC). Stationary effects, such as tensor light shift \cite{Beato_Palacios-Laloy_2020, Happer_Mathur_1967}, can also lead to the AOC due to symmetry violation. Orientation, in turn, can cause circular birefringence and, as a consequence, rotation of the plane of polarization of the pump light. 

There are several arguments against orientation dominance in our experiment: firstly, the signals we observed were recorded with the laser tuned to the center of the resonance line \textit{F}~=~3 > \textit{F'}~=~3 transition of $D_{1}$ absorption line, which provides very poor conditions for observing circular birefringence. But in complex overlapping spectra, the signal of circular birefringence is observed almost along the entire absorption line profile, and the amplitude of the orientation signals is usually an order of magnitude or two larger than the amplitude of the alignment signals. 

Secondly, the signals in Figure~\ref{figure4} show a clear symmetric dependence on the magnetic field direction in the 0\textit{xy} plane, which should not be the case in the case of orientation. But in principle a similar, although not so strong, dependence can be due to the broadening effect of the linearly polarized component of the light \cite{Petrenko_Pazgalev_Vershovskii_2021}. 

Thirdly, the very fact that the longitudinal with respect to the beam field $\bf{B}_{\tt{z}}$ causes Hanle resonances, and the sign of these resonances changes to the opposite when the direction of $\bf{B}_{\tt{z}}$ changes (Figures \ref{figureB3}, \ref{figureB4}) confirms that these are alignment signals: in the case of orientation, Hanle resonances do not appear at all when the offset field is directed along the \textit{z} axis.

In order to dispel any remaining doubts, we conducted a control experiment: using the additional quarter-wave plate, we introduced some ellipticity into the light (Figure~\ref{figure6}) and showed that introducing right and left circular polarization leads to the transformation of a symmetric signal into antisymmetric ones. The polarity of these antisymmetric signals changes to the opposite both when the direction of circular polarization changes and when the direction of the magnetic field changes, as should be the case in the case of orientation.  The maximum amplitude of orientation signals exceeds the amplitude of the alignment signal by about 40 times. Note the small width of the symmetric (presumably pure alignment) signal compared to the pure orientation signals at constant pump intensity. 

Figure~\ref{figure7} shows a comparison of the resonance widths measured in zero field with theoretical estimates for orientation \cite{Seltzer_2008}. It should be noted that the contribution to the broadening from collisions with the cell walls is the same for all moments and, unlike other contributions to $T^{(1)}_{1}$, is not subject to changes introduced by the nuclear slow-down factor. The recalculation of the relaxation rates we measured into units of frequency was carried out taking into account the change in the effective gyromagnetic ratio in low fields \cite{Seltzer_2008}.

The red line, which roughly corresponds well to the experimental points, represents the width (HWHM) of the \textit{orientation} resonances calculated for a field of 65~nT. Despite the discrepancy (65~nT vs zero), it can be seen that the relaxation rates we observed do not fall into the ``forbidden'' zone, and are clearly in the SERF zone. For a complete comparison, it should be recalled that under identical conditions, the width of the \textit{alignment} resonance is half the width of the orientation resonance ($\Delta^{(2)} \approx 1/(2T^{(2)}_{2})$ \cite{Breschi_Weis_2012,Meraki_Elson_Ho_Akbar_Kozbial_Kolodynski_Jensen_2023}.

Thus, we have some reason to claim that the SERF effect was demonstrated in alignment. 

However, as noted in the Introduction, conservation of the alignment requires a symmetry of a higher order than the standard space-time symmetries. Taking this into account, a less provocative interpretation of the results obtained can be proposed -- e.g., that in the zero field, as a result of collective effects (namely, fast spin exchange in the absence of precession), a \textit{local} spontaneous transformation of alignment into orientation occurs, similar to the formation of domains in a ferromagnet below the Curie point. In this state, the total zero moment of the whole ensemble could be ensured if the sum of the moments of the domains is equal to zero, the orientation within each domain could be maintained by fast spin exchange in accordance with the law of conservation of angular momentum, and for an observer the state of the entire ensemble would look like alignment.

In any case, the demonstrated effect, in our opinion, deserves further study.

\section{Conclusions}\label{sec:6}  

We have shown that transient processes occurring in the alignment under pumping with linearly polarized light and under switching of transverse magnetic fields are, under certain conditions, determined by longitudinal relaxation with a characteristic time $T^{(2)}_{1}$, and spin-exchange interactions do not contribute to $T^{(2)}_{1}$. We have also shown that, at sufficiently high temperatures and pump intensities, the contribution of spin-exchange broadening to the resonance width ($\Delta^{(2)} \approx 1/(2T^{(2)}_{2})$) can be reduced practically to zero, much like in SERF sensors. This effect deserves further study, since the laws of conservation of angular momentum that determine the SERF effect in orientation should not guarantee a similar effect in the alignment.

Indeed, in a bias magnetic field whose vector is directed along the vector of the electric component of light, the alignment relaxes significantly faster than the orientation should relax. This, however, does not happen in the zero field, nor in the field applied in the perpendicular direction.

The results obtained are interesting from a theoretical point of view, and we hope that they will attract the attention of theoreticians. At the same time, they have specific practical significance, since they demonstrate the fundamental possibility of creating a compact SERF sensor based on the alignment effect in a single-beam scheme with a linearly polarized pump-detection beam and with balanced detection of the polarization rotation signal.

\textbf{}

\textbf{Conflict of interest:} The authors declare that they have no conflict of interest.

\textbf{}

\textbf{Funding:} This research was funded by the baseline project FFUG-2024-0039 at the Ioffe Institute.

\textbf{}

\textbf{Acknowledgments:} The authors thank Prof.~Eugene B.~Aleksandrov and Dr.~Anatoly S.~Pazgalev for valuable discussions.
\textbf{}

\renewcommand{\thefigure}{B\arabic{figure}}
\setcounter{figure}{0}

\renewcommand{\theequation}{A\arabic{equation}}
\setcounter{equation}{0}

\section*{Appendix A. Stationary balanced signal} 
\addcontentsline{toc}{section}{Appendix}

\begin{figure}[!t]  
	\includegraphics[width=\linewidth]{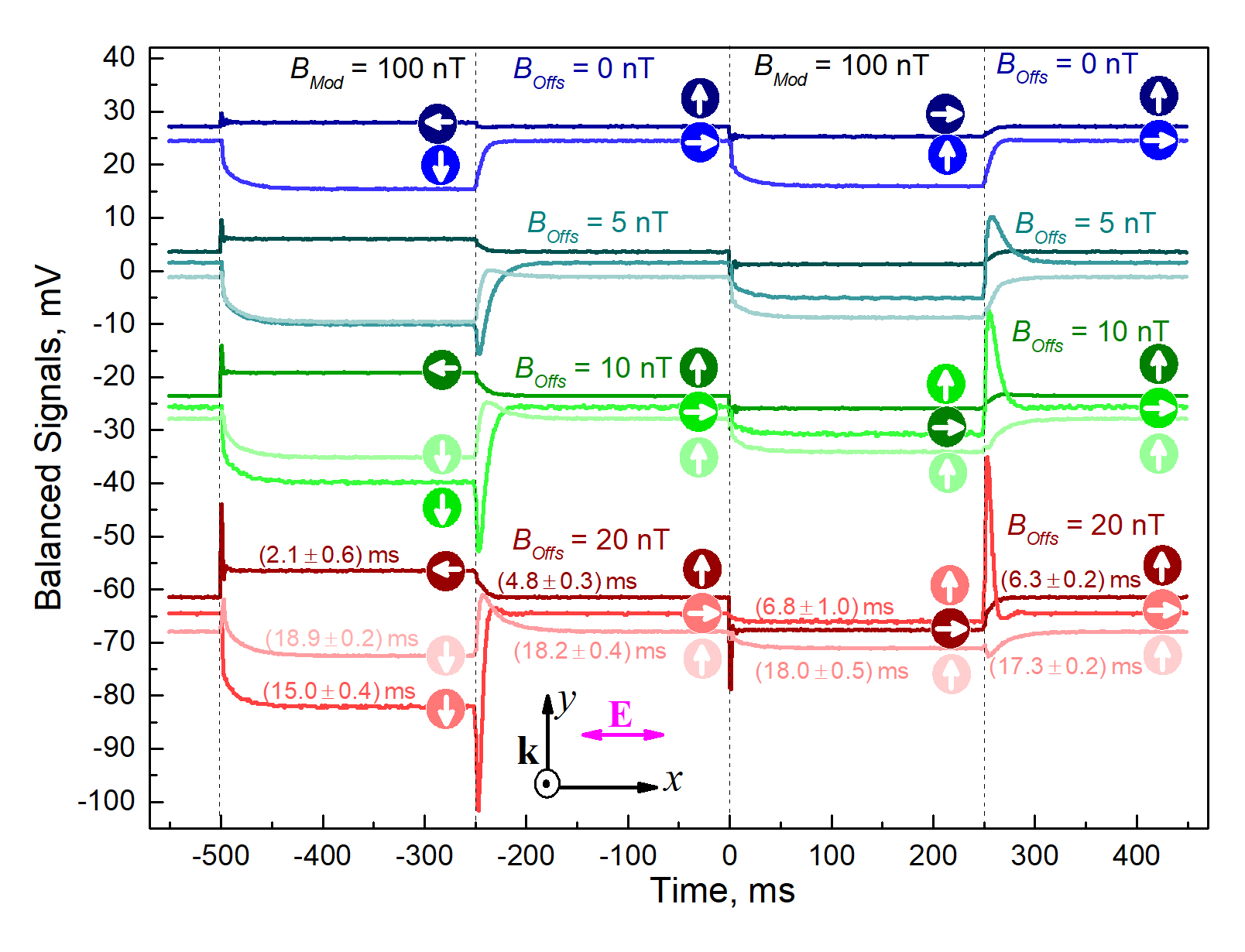}
	\caption{Examples of recordings of polarization rotation signals when switching the induction and direction of the magnetic field. The vertical shift of the graphs is arbitrary. Arrows designate the direction of the magnetic field in the 0\textit{xy} plane (polarization vector \textbf{E}\textbar \textbar \textbf{x}, wave vector \textbf{k}\textbar \textbar \textbf{z}) in a given recording interval, the numbers above the arrows are the modulus of the magnetic field induction. Several values of the relaxation times are given.}\label{figureB1}
 
\end{figure} 
\begin{figure}[!t]  
	\includegraphics[width=\linewidth]{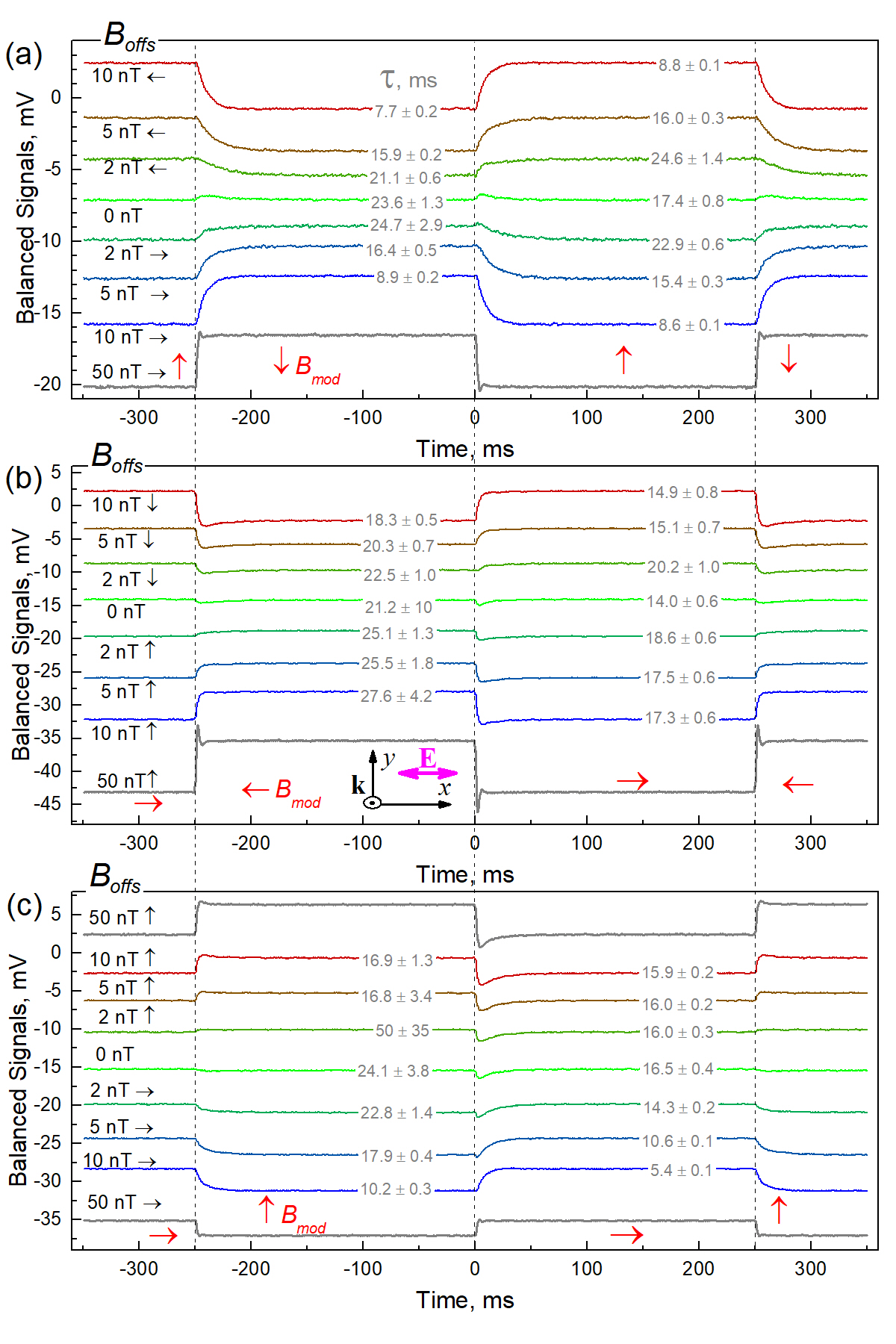}
	\caption{Examples of signal recordings made during magnetic field switching (\textit{B}${}_{Mod}$ = 5~nT). Arrows indicate the direction of the modulation magnetic field \textbf{B${}_{\tt{Mod}}$} (red) and the DC bias field \textbf{B${}_{\tt{Offs}}$} (gray) in the 0\textit{xy} plane in a given recording interval; the direction of \textbf{E} is considered horizontal (\textbf{E}\textbar \textbar \textbf{x}, \textbf{k}\textbar \textbar \textbf{z}). The values of relaxation times are given on the corresponding lines.}\label{figureB2}
  
\end{figure}

In the \textit{x'y'z} coordinate system related to the direction of the magnetic field, the passage of the probe beam with the polarization azimuth \textit{$\varphi $} through a medium is described using the Jones formalism as follows:
\begin{equation}
    \left(\begin{array}{c} {E_{Lx'} } \\ {E_{Ly'} } \end{array}\right)=\mathbf{T}\cdot \left(\begin{array}{c} {E_{x'} } \\ {E_{y'} } \end{array}\right)
    \label{eq:equat_A1}
\end{equation}
where $E_{x}=E \cdot Cos(\varphi)$, $E_{y}=E \cdot Sin(\varphi)$, $E$ is the amplitude of the electric component of the light wave, \textbf{T} is the Jones matrix containing real coefficients $T_{ij}$ (which corresponds to the absence of birefringence):
\begin{equation}
    \mathbf{T}=\left(\begin{array}{cc} {T_{x'x'} } & {0} \\ {0} & {T_{y'y'} } \end{array}\right),
    \label{eq:equat_A2}
\end{equation}
where the off-diagonal elements \textit{T${}_{x'y'}$} and \textit{T${}_{y'x'}$} are equal to zero. \textbf{I${}_{0}$} is the probe intensity vector:
\begin{equation}
   \mathbf{I}_{0} =\left(\begin{array}{c} {I_{x'} } \\ {I_{y'} } \end{array}\right)=\left(\begin{array}{c} {I_{0} \cdot \cos ^{2} \left(\varphi \right)} \\ {I_{0} \cdot \sin ^{2} \left(\varphi \right)} \end{array}\right)
    \label{eq:equat_A3}
\end{equation}

Let us define dichroism as the relative difference in transmission coefficients:
\begin{equation}
  \Delta T\equiv \frac{T_{xx}^{2} -T_{yy}^{2} }{2} ,T_{Mean} \equiv \frac{T_{xx}^{2} +T_{yy}^{2} }{2} , 
    \label{eq:equat_A4}
\end{equation}

\begin{figure}[!t]  
	\includegraphics[width=\linewidth]{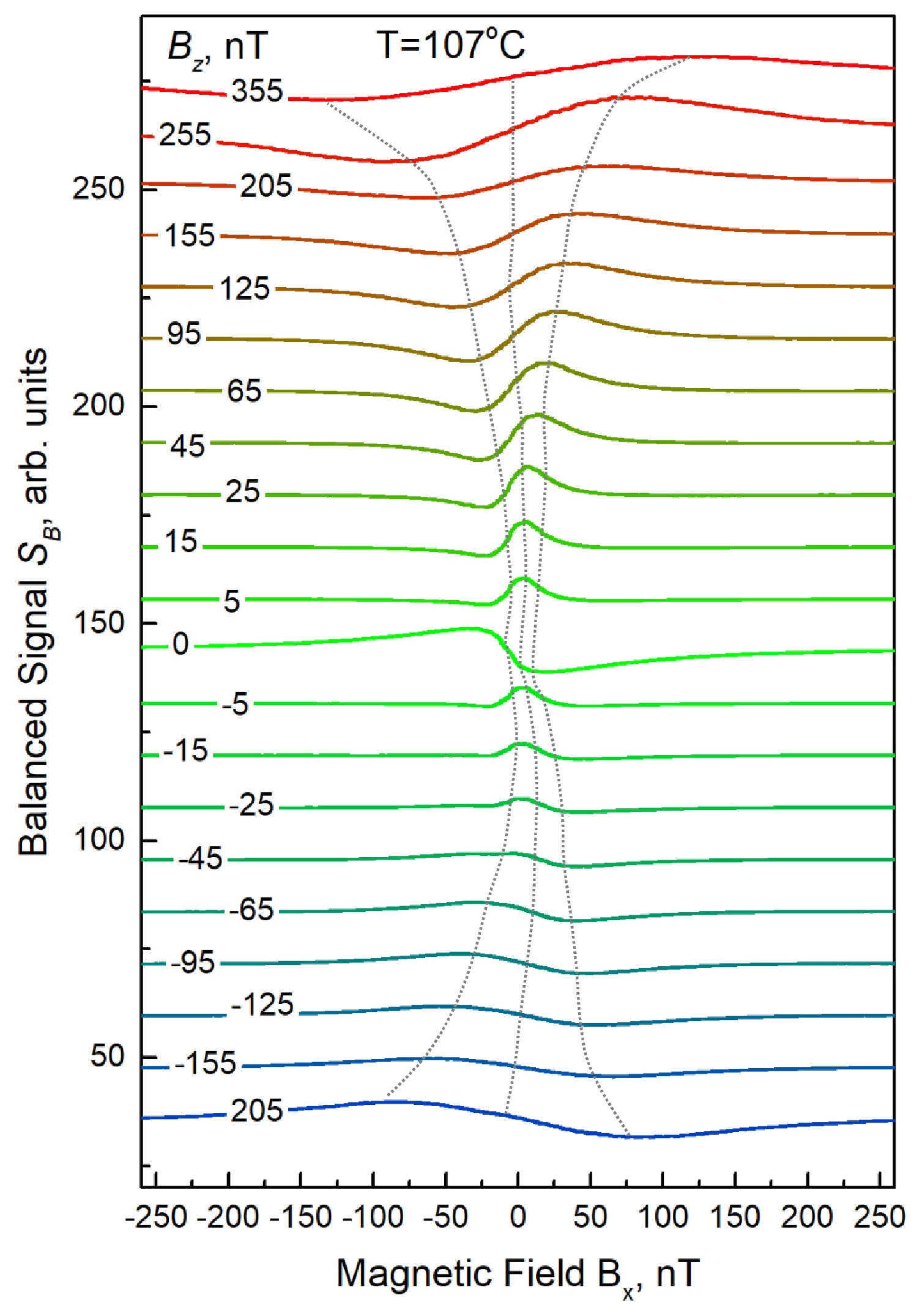}
	\caption{Examples of demodulated signals recorded under modulation and scanning of the field \textit{B${}_{x}$} at the different values of the field \textit{B${}_{z}$} at T~=~107~${}^{o}$C (see Figure~\ref{figure5}). The vertical scale for each record is chosen to provide approximately the same span; the actual amplitude data are shown in Figure~\ref{figure5}c. The dotted gray lines indicate approximately the position of the center of the resonance line and the midpoints of its slopes.}\label{figureB3}  
\end{figure} 

\begin{figure}[!t]  
	\includegraphics[width=\linewidth]{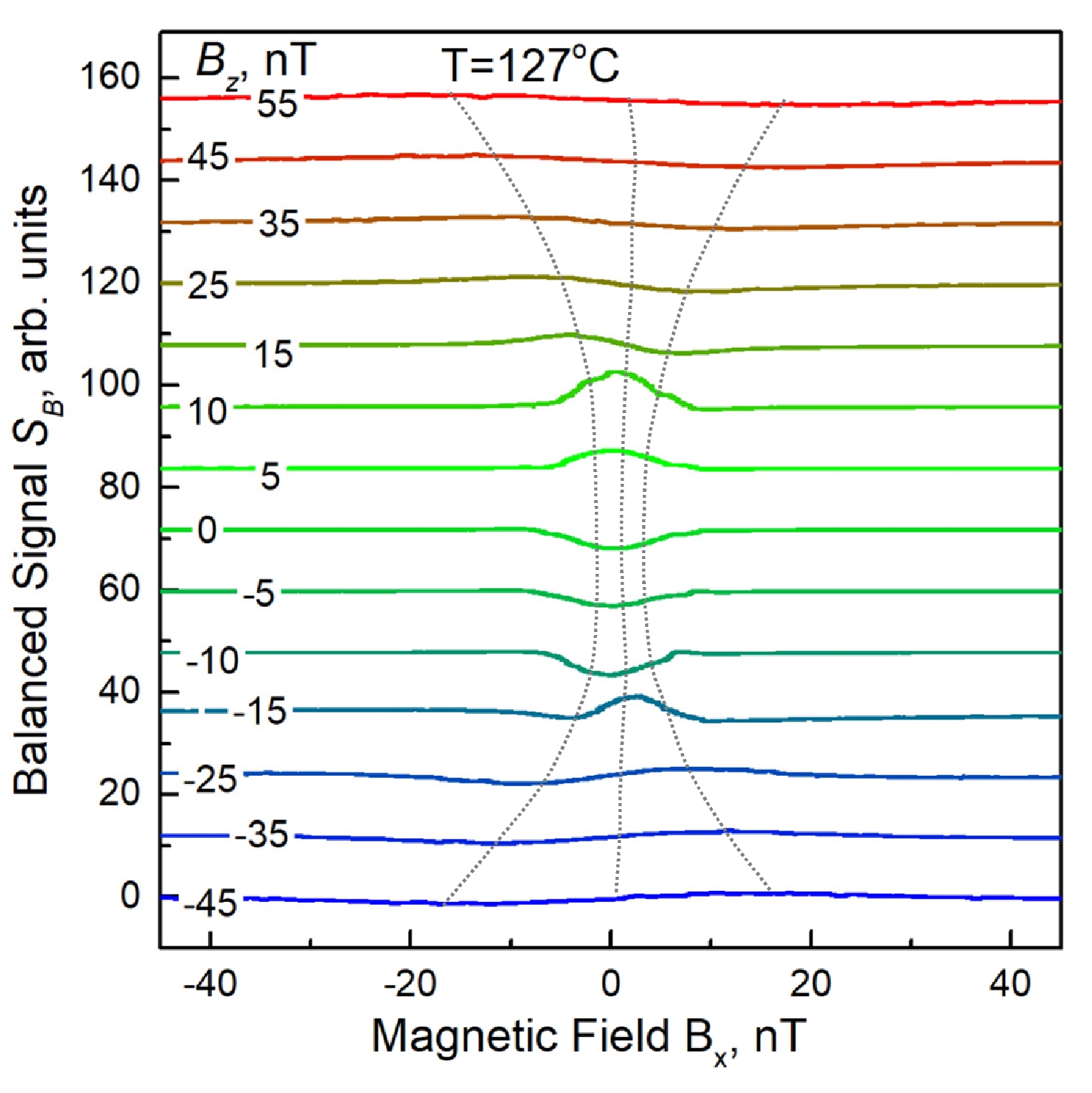}
	\caption{Examples of demodulated signals recorded under modulation and scanning of the field \textit{B${}_{x}$} at the different values of the field \textit{B${}_{z}$} at T~=~127~${}^{o}$C (see Figure~\ref{figure5}). The vertical scale for each record is chosen to provide approximately the same span; the actual amplitude data are shown in Figure~\ref{figure5}c. The dotted gray lines indicate approximately the position of the center of the resonance line and the midpoints of its slopes.}\label{figureB4}   
\end{figure}

\textit{$\Delta $T} is proportional to $\rho^{(2)}_{0}$. The total \textit{S${}_{T}$} signal measured in transmitted light is simply equal to the intensity at the cell output. The magnitude of the \textit{S${}_{B}$} signal, equal to the difference in intensities in two perpendicular polarizations, is determined by the angle \textit{$\Delta $$\varphi $}~=~\textit{$\varphi $--$\varphi $${}_{0}$ }of rotation of the beam polarization plane signal, 
\begin{equation}
 S_{B} =-\frac{I_{Out} }{2} \sin (2\Delta \varphi )\approx -I_{Out} \cdot \Delta \varphi .  
    \label{eq:equat_A5}
\end{equation}
where \textit{I${}_{Out}$} is the light intensity at the sell output. At $\Delta \varphi \ll 1$, which is always the case in the case of linear dichroism signals, $S_{B} \sim \Delta \varphi$. The theoretical value of the \textit{S${}_{B}$} signal can be calculated using expressions Eqs. (\ref{eq:equat_A1}), (\ref{eq:equat_A2}) followed by translating to the\textit{ xyz} coordinate system related to the beam polarization direction (\textbf{E}\textbar \textbar \textbf{x}, \textit{$\varphi $} = arctan(\textit{B${}_{y}$/B${}_{x}$})), and then rotating the coordinate system by $\pi $/4 and calculating the difference in the intensities of the corresponding polarization components:
\begin{equation}
  \left(\begin{array}{c} {E_{Rx''} } \\ {E_{Ry''} } \end{array}\right)=\left(\begin{array}{cc} {\frac{\sqrt{2} }{2} } & {-\frac{\sqrt{2} }{2} } \\ {\frac{\sqrt{2} }{2} } & {\frac{\sqrt{2} }{2} } \end{array}\right)\cdot \left(\begin{array}{c} {E_{Lx} } \\ {E_{Ly} } \end{array}\right), 
    \label{eq:equat_A6}
\end{equation}
and
\begin{equation}
  S_{B} =E_{Ry''}^{*} \cdot E_{Ry''} -E_{Rx''}^{*} \cdot E_{Rx''} 
    \label{eq:equat_A7}
\end{equation}

Thus, for signal \textit{S${}_{B}$} for $\alpha $~=~\textit{$\Delta $T/T${}_{Mean}$}~$<$$<$~1 we obtain
\begin{equation}
  S_{B} \approx \Delta T\cdot \left(Sin(2\varphi )+\frac{1}{4} \alpha \cdot Sin(4\varphi )\right). 
    \label{eq:equat_A8}
\end{equation}
Consequently, \textit{S${}_{B}$}~=~0 both when \textbf{B}\textbar \textbar \textbf{x} (\textit{$\varphi $}~=~0) and when \textbf{B}\textbar \textbar \textbf{y }(\textit{$\varphi $}~=~$\pm$$\pi $/2).\textbf{}

\section*{Appendix B. Examples of raw recordings} 
\addcontentsline{toc}{section}{Appendix}

In Figure~\ref{figureB1} we present recordings of transient processes taken in various configurations and the results of their processing.

In Figure~\ref{figureB2} we present raw data for Figure~\ref{figure3}b -- the recordings of transient processes taken in various configurations and the results of their processing.

In Figures \ref{figureB3}, \ref{figureB4}  we present raw data for Figure~\ref{figure4}  -- the examples of demodulated signals recorded under modulation and scanning of the field \textit{B${}_{x}$} at the different values of the field \textit{B${}_{z}$} at T~=~107~${}^{o}$C and T~=~127~${}^{o}$C.


%

\end{document}